\documentclass[12pt,preprint]{emulateapj}
\usepackage{graphicx}
\usepackage{natbib}
\shorttitle{On The Origin Of High Energy Correlations in Gamma-ray Bursts}
\shortauthors{D. Kocevski}

\begin{document}

\title{On The Origin Of High Energy Correlations in Gamma-ray Bursts}

\author{Daniel Kocevski \altaffilmark{1}}

\altaffiltext{1}{Kavli Institute for Particle Astrophysics and Cosmology, Stanford University, 2575 Sand Hill Road M/S 29, Menlo Park, Ca 94025 }


\begin{abstract}

I investigate the origin of the observed correlation between a gamma-ray burst's ${\nu}F_{\nu}$ spectral peak $E_{\rm pk}$ and its isotropic equivalent energy $E_{\rm iso}$ through the use of a population synthesis code to model the prompt gamma-ray emission from GRBs.  By using prescriptions for the distribution of prompt spectral parameters as well as the population's luminosity function and co-moving rate density, I generate a simulated population of GRBs and examine how bursts of varying spectral properties and redshift would appear to a gamma-ray detector here on Earth. I find that a strong observed correlation can be produced between the source frame $E_{\rm pk}$ and $E_{\rm iso}$ for the detected population despite the existence of only a weak and broad correlation in the original simulated population.  The energy dependance of a gamma-ray detector's flux-limited detection threshold acts to produce a correlation between the source frame $E_{\rm pk}$ and $E_{\rm iso}$ for low luminosity GRBs, producing the left boundary of the observed correlation.  Conversely, very luminous GRBs are found at higher redshifts than their low luminosity counterparts due to the standard Malquest bias, causing bursts in the low $E_{\rm pk}$, high $E_{\rm iso}$ regime to go undetected because their $E_{\rm pk}$ values would be redshifted to energies at which most gamma-ray detectors become less sensitive.  I argue that it is this previously unexamined effect which produces the right boundary of the observed correlation.  Therefore, the origin of the observed correlation is a complex combination of the instrument's detection threshold, the intrinsic cutoff in the GRB luminosity function, and the broad range of redshifts over which GRBs are detected. Although the GRB model presented here is a very simplified representation of the complex nature of GRBs, these simulations serve to demonstrate how selection effects caused by a combination of instrumental sensitivity and the cosmological nature of an astrophysical population can act to produce an artificially strong correlation between observed properties.

\end{abstract}


\section{Introduction}

The search for empirical correlations between observable parameters has long been an important path to understanding the underlying nature of astrophysical sources.  Such attempts at correlating data, though, are inherently risky, as without careful consideration of observational biases, one can be lead to false conclusions regarding the strength and nature of observed correlations.  In this paper I consider the origin of the much discussed correlation between a gamma-ray burst's ${\nu}F_{\nu}$ spectral peak ($E_{\rm pk}$) and its isotropic equivalent energy ($E_{\rm iso}$), first reported by \citet{Amati02}.  The importance of such a correlation cannot be understated, as a tight relationship between a GRB's spectral properties and total energetics would allow for the distance of cosmological GRBs to be determined from gamma-ray data alone, opening the possibility of using high redshift GRBs for cosmological distance-scale applications.

Discussions regarding the origin of the $E_{\rm pk} - E_{\rm iso}$ correlation have had a long and contentious history in the literature.  \citet{Nakar05} and \citet{Band05} showed that a significant fraction of BATSE detected GRBs without known redshift could not be consistent with the relation at any redshift, indicating that the true correlation may in fact be much broader than the one originally found by \citet{Amati02}.  Likewise, \citet{Butler07, Butler09} showed that hard and under-luminous GRBs detected by the \emph{Swift} spacecraft's Burst Alert Telescope (BAT) \citep{Gehrels04} were inconsistent with the pre-\emph{Swift} $E_{\rm pk} - E_{\rm iso}$ correlation.  Furthermore, the normalization of the \emph{Swift} based correlation shifts towards the detection threshold of the BAT, suggesting that the strength of the correlation is largely governed by the sensitivity of the collecting instrument.   

Questions regarding the influence of detector thresholds on the $E_{\rm pk} - E_{\rm iso}$ correlation were examined in detail for a variety of instruments by \citet{Ghirlanda08} and \citet{Nava08}.  In \citet{Nava08}, the authors claim that evidence for the effects of detections thresholds are clearly present in their spectroscopic samples, but that the strength and shape of the resulting correlation is not governed by these effects.  

More recently, \citet{Butler10} applied a multi-variant analysis to over 200 \emph{Swift} detected GRBs in order to investigate their intrinsic energetics distribution and spectral parameters.  The authors found evidence for a very broad correlation between the source frame $E_{\rm pk}$ and $E_{\rm iso}$, which only takes the form observed by \citet{Amati02} through the non-detection of weak events, matching the conclusions drawn by \citet{Nakar05, Band05}.

In this paper, I examine the nature of the $E_{\rm pk} - E_{\rm iso}$ correlation through a new approach that utilizes a population synthesis code to model the prompt gamma-ray emission from GRBs in order to examine how bursts of varying spectral properties and redshift would appear to a gamma-ray detector here on Earth.  This work is similar to the analysis performed by \citet{Nava08} in that I try to quantify the degree to which detector thresholds produce the observed $E_{\rm pk} - E_{\rm iso}$ correlation, although I take the additional step of using prescriptions for the population's luminosity function and co-moving rate density to examine how the cosmological distribution of these events biases the observed properties of the detected population.

I find that a strong observed correlation can be produced between the source frame $E_{\rm pk}$ and $E_{\rm iso}$ for the detected population despite only a week and broad correlation being present in the original simulated population.  The origin of this observed correlation is a complex combination of the instrument's detection threshold, the intrinsic cutoff in the GRB luminosity function, and the broad range of redshifts over which GRBs are detected.  

Any energy dependance on the flux limited detection threshold of a gamma-ray detector acts to produce a correlation between the inferred source frame $E_{\rm pk}$ and $E_{\rm iso}$ for low luminosity GRBs, producing the left boundary of the observed  $E_{\rm pk} - E_{\rm iso}$ correlation.  Conversely, very luminous GRBs are relatively rare and therefore probability dictates that we would have to look to higher redshifts in order to detect these events. Therefore, GRBs in the low $E_{\rm pk}$, high $E_{\rm iso}$ regime would go undetected because their $E_{\rm pk}$ values would be redshifted to energies at which most gamma-ray instruments become less sensitive, producing the right boundary of the observed correlation.  The combination of these selection effects act to produce a relatively tight correlation between $E_{\rm pk}$ and $E_{\rm iso}$ in the population detected by the instrument, despite the existence of a much broader intrinsic correlation that goes unseen by the observer.

I present an overview of the population synthesis code used in this paper in Section 2.1, followed by a more in-depth description of the code in Sections 2.2, 2.3, and 2.4.  I present the burst demographics of a simulation of 40,000 GRBs in Section 3 and discuss the implications of these results in regards to the observed $E_{\rm pk} - E_{\rm iso}$ correlation in Section 4.  Throughout the paper, I will refer to the source and observer frame $\nu F_{\nu}$ peak as $E_{\rm pk}$ and $E_{\rm pk,obs}$ respectively.

\section{GRB Model} \label{sec:Models}

\subsection{Model Overview} \label{sec:ModelOverview}

Gamma-ray burst continuum spectra can evolve quite dramatically over the course of a burst.  This evolution is generally characterized by an overall softening of the spectra to lower energies, with the peak of the ${\nu}F_{\nu}$ spectrum evolving through the detector bandpass over the duration of the burst.  This evolution will be delayed by the effects of time dilation for GRBs at high redshifts, resulting in a longer observed spectral lag between the high and low energy channels and a broadening of the pulse profile.    At the same time, the observed GRB flux falls as a function of increasing luminosity distance.  The net effect is that soft and faint emission will become increasingly difficult to observe with gamma-ray detectors such as \emph{Swift} or \emph{Fermi's} Gamma-ray Burst Monitor (GBM) \citep{Meegan09}, which suffer large drops in sensitivity at energies below $\sim$20 KeV.  The observed bursts properties are therefore a complex convolution of the effects of cosmological redshift and detector sensitivity and hence I turn to simulations to obtain a better idea of how these bursts would appear in the observer frame.  

Two empirical correlations form the basis for the GRB model that I have developed to investigate this question.  The first is the {\it hardness-intensity correlation} or HIC, which relates the instantaneous hardness of the spectra and the instantaneous energy flux $F_{E}$, within individual pulses. For the decay phase of a pulse, the most common behavior of the HIC is a power-law relationship between $F_{E}$ and the peak of the ${\nu}F_{\nu}$ spectrum, $E_{\rm pk,obs}$, of the form  $F_{E}\propto E_{\rm pk,obs}^{\eta}$ \noindent, where $\eta$ is the HIC power-law index. The second correlation is the {\it hardness-fluence correlation} or HFC (Liang $\&$ Kargatis 1996) which describes the observation that the instantaneous hardness, or $E_{\rm pk,obs}$, of the spectra decays exponentially as a function of the time-integrated flux, or fluence, of the burst.  The HFC can be stated as $E_{\rm pk} = E_{0} e^{-\Phi / \Phi_{0}}$, where $\Phi(t)$ is the photon fluence integrated from the start of the burst and $\Phi_{0}$ is the exponential decay constant.  

\citet{Kocevski03} show that both the HIC and HFC correlations can be produced through simple relativistic kinematics when applied to a spherical shell expanding at relativistic velocity.  The curvature of a relativistic fireball would make the photons emitted off the line of sight delayed and affected by a varying Doppler boost due to the increasing angle at which the photons were emitted with respect to the observer. Simulating a GRB pulse using only these two correlations to describe the evolution of a Band photon spectrum  \citep{Band93} as a function of time reproduces the Fast Rise Exponential Decay (FRED) pulse shape that is so ubiquitous in GRB data. 

With this time-resolved spectral model, I can simulate a GRB at a variety of redshifts and quantify the effects that redshift have on the inferred pulse properties.  In order to convert our modeled photon spectrum into a count spectrum and eventually a count light curve, I take the simulated photon spectrum and fold it through an instrument response matrix.  I then add a Poisson distributed, energy dependent, background spectrum derived from the median backgrounds of a sample of detected bursts.  From this count+background spectrum, I can then calculate the resulting count light curve that the instrument would have produced for a given input pulse model.

Using the simulated count light curve and spectra, I can then back out the burst's trigger significance, duration, hardness, peak flux, and time integrated spectral parameters as would be inferred by the observer.  Using the known redshift, I can then use these properties to determine the inferred source frame $E_{\rm pk}$ and obtain an estimate of $L_{\rm iso}$, and $E_{\rm iso}$.  By generating a large number of GRBs following a realistic redshift distribution, I can examine the distribution of properties for the bursts that are detected by the instrument and compare them to the true source frame distributions known otherwise only to nature.  

Sections 2.2, 2.3, and 2.4 describe the model outlined above in more detail.  The model is currently written in a combination of IDL and Python and is available upon request\footnote{http://www.kocevski.com/GRBModel/}.

\subsection{Modeling Individual FRED Pulses} \label{sec:FREDs}

In order to model a single FRED pulse as it would appear in the source frame of the GRB, I start by assuming an empirical Band photon spectrum and evolving its $E_{\rm pk}$ and peak energy luminosity $L_{\rm E}$ at $E_{\rm pk}$ through time.  Drawing upon the extensive BATSE spectral catalog presented by \citep{Preece00}, I set the low and high energy power-law indices of the simulated spectrum to match the median BATSE determined values of $\alpha_{\rm pk} = -1.1$ and $\beta_{\rm pk} = -2.3$ respectively.  Likewise, I draw our initial $E_{\rm pk,0}$ from a log-normal distribution centered at $E_{\rm pk,obs} \sim 200 \times (1+{\bar z})$ keV and falling off sharply above $300 \times (1+{\bar z})$ keV.  Here, ${\bar z}$ represents the redshift at which we expect to see the largest number of events, assumed here to be the redshift at which the cosmic star formation peaks.  

I then evolve the spectrum's $E_{\rm pk}$ and $L_{\rm E}$ in time following the relations derived by \citet{Kocevski03} for the effects of a spherical shell expanding at relativistic velocity, namely:

\begin{equation} \label{Eq:Epkt}
E_{\rm pk}(t)=E_{\rm pk,0}{\cal D}=\frac{E_{\rm pk,0}}{(1+t/\tau_{\rm ang})}
\end{equation}
\begin{equation} \label{Eq:Ft}
L_{\rm E}(t) = L_{\rm E,0} {\cal D}^{2} = \frac{L_{\rm E,0}}{(1+t/\tau_{\rm ang})^2}
\end{equation}
where $E_{\rm pk,0}$ and $L_{\rm E,0}$ represent the initial $E_{\rm pk}$ and $L_{\rm E}$ values in the emitting surface's co-moving frame.  ${\cal D}$ is the angle dependent Lorentz boosting factor for transformations from this co-moving frame to the GRB's source frame (i.e the rest frame of the progenitor), given by:

\begin{equation} \label{doppler}
{\cal D} (\Gamma,\mu)=\frac{1}{\Gamma (1-\beta \mu)} =\Gamma
(1+\beta \mu'), \label{boost}
\end{equation}

\begin{figure}
\includegraphics[height=.23\textheight,keepaspectratio=true]{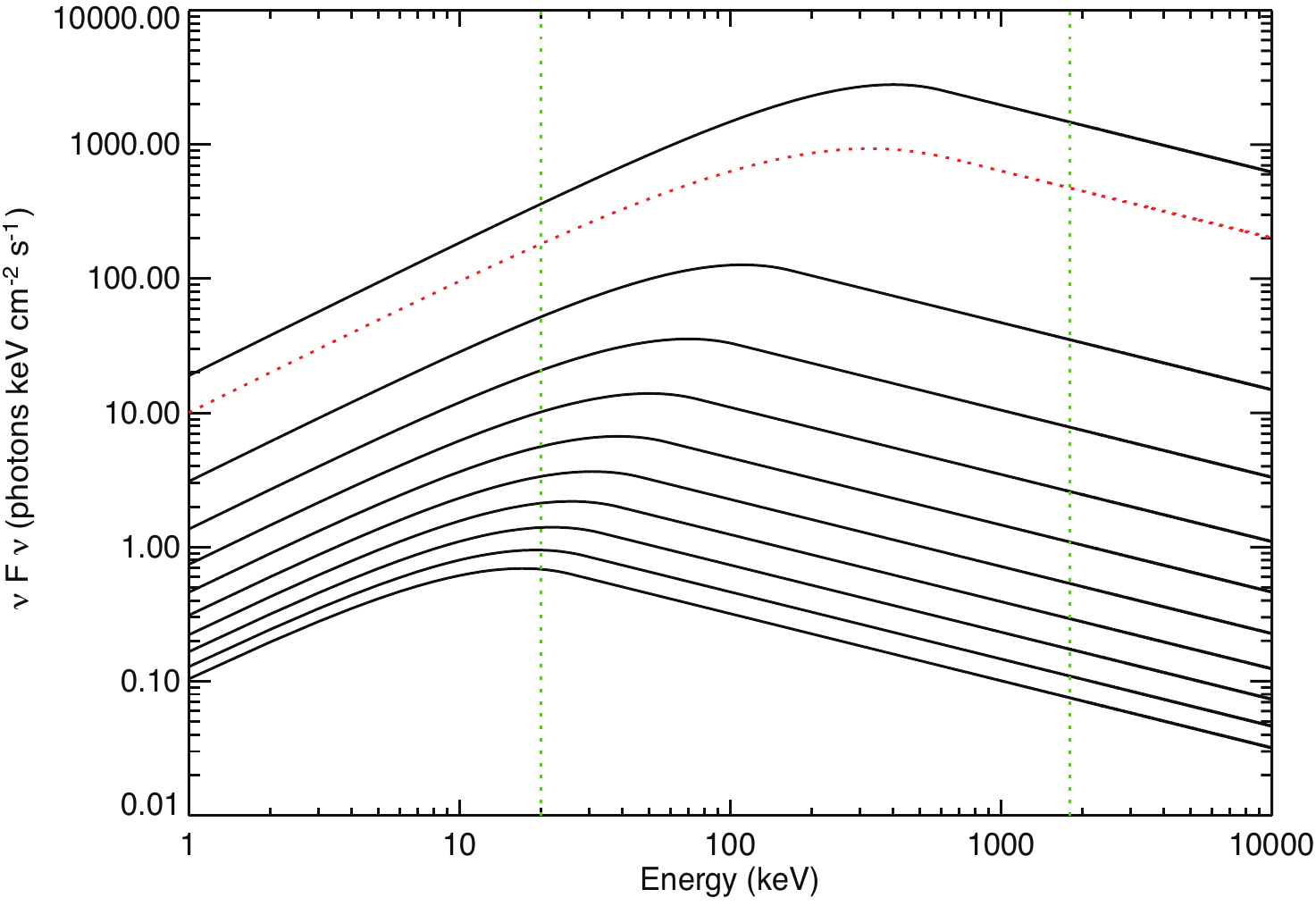}
\caption{The hard to soft spectral evolution of $E_{\rm pk}$ as a function of time as prescribed by relativistic kinematics for a spherical shell traveling towards the observer with high Lorentz factor.}
\label{SpectralEvolution}
\end{figure}

The combined effect of the angle dependent Lorentz boost factor and the additional time of flight of photons emitted off the line of sight is the delayed arrival of soft photons.  Even in the idealized case of a curved, relativistic emitting surface, which emits for an instantaneous moment in time and at a single frequency $E_{\rm pk,0}$ with luminosity $L_{\rm E,0}$, the observed profile will be a broad pulse with harder emission arriving first followed by the arrival of softer and weaker emission. Equations \ref{Eq:Epkt} and \ref{Eq:Ft} only describe the decay phase of the pulse, so I take the additional step of introducing a power-law rise term to the analytic function describing peak energy flux such that $L_{\rm E}(t) = L_{\rm E}(t) \times (t/t_{0})^{r}$.  Varying the $t_{0}$ for this power law component can adjust the resulting time to peak flux accordingly, allowing for the production of a FRED pulses with a variety of durations.  

\begin{figure} 
\includegraphics[height=.23\textheight,keepaspectratio=true]{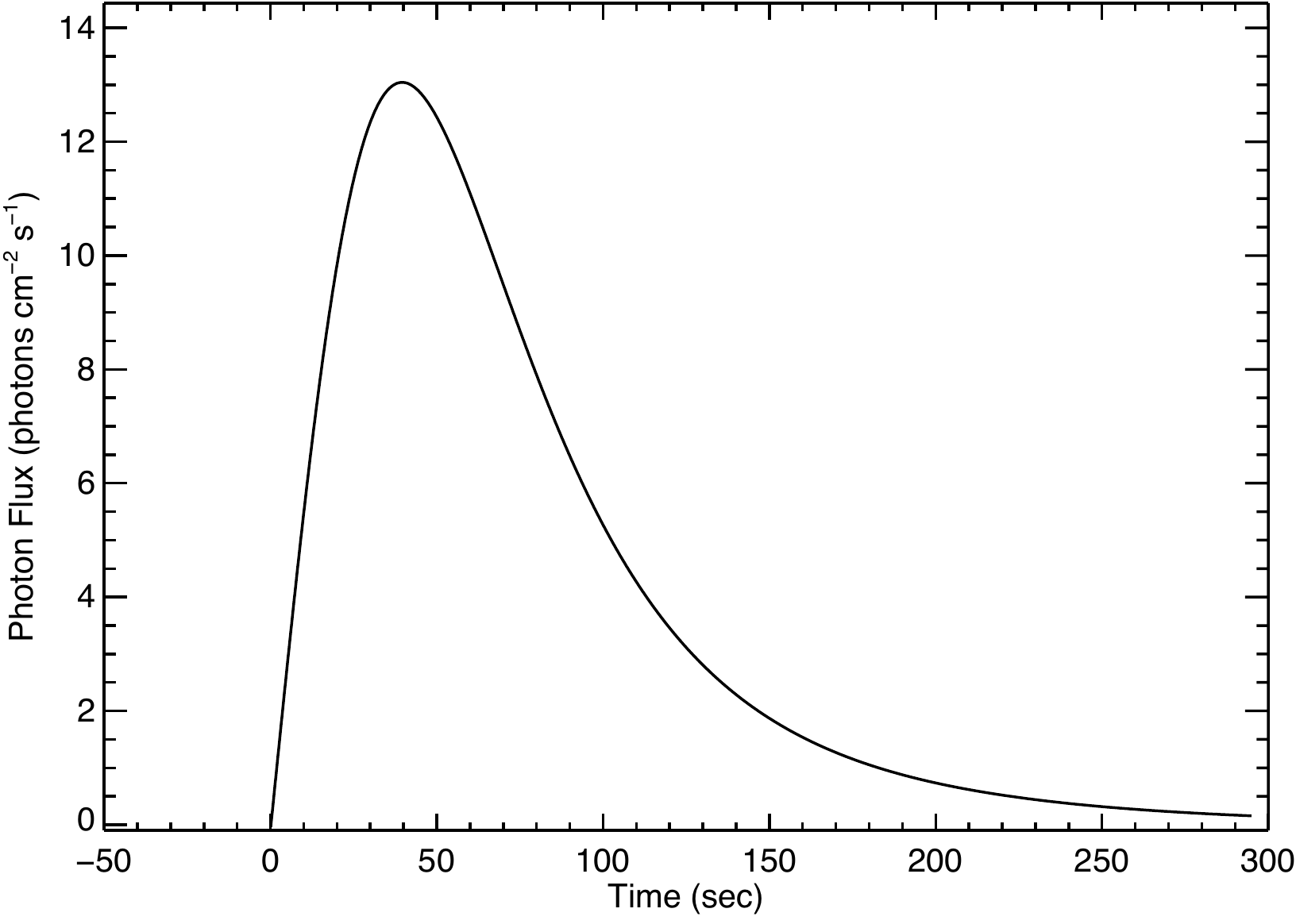}
\caption{The resulting fast rise exponential decay (FRED) pulse shape generated by the model by integrating the bolometric GRB spectrum at each time bin.}
\label{PhotonFluxLightCurve}
\end{figure}

An example of the hard to soft evolution that results from Equations \ref{Eq:Epkt} and \ref{Eq:Ft} can be seen in Figure \ref{SpectralEvolution}.  Both $E_{\rm pk}$ and $L_{\rm E}$ decrease with time, with the red dotted line representing the time integrated spectrum produced by summing the individual time-resolved spectra.  The green dotted lines represent the nominal energy window of the BATSE instrument.  A 3-dimensional array is used to store the time-resolved photon spectra calculated at one second intervals, producing a photon data cube as a function of energy, time, and photon flux.  An example of a bolometric photon light curve, produced by integrating each time-resolved photon spectra over energy, can be see in Figure \ref{PhotonFluxLightCurve}.  The resulting light curve reproduces the ubiquitous fast rise exponential decay profiles observed in GRB time histories.  

\subsection{Population Synthesis} \label{sec:Population}

The cosmological distances at which GRBs occur play a non-trivial role in the resulting properties seen in the observer frame.  As the redshift of a GRB increases, $E_{\rm pk,obs}$ is pushed closer to the lower bound of the detector's energy window.  As this happens, more of the GRB's low energy spectrum goes undetected by the instrument, reducing the observed flux and fluence beyond what is expected from simply increasing the GRB's luminosity distance.  Therefore, it is important to simulate a population of GRBs with a realistic redshift distribution in order to properly account for this effect.


To do so, I turn to the luminosity function $\phi(z)$ and co-moving rate density $\dot{\rho}(z)$ estimates presented in \citet{Butler10}, where the authors utilize a multi-variant analysis on over 200 \emph{Swift} detected GRBs with known redshift to infer $\phi(z)$ and $\dot{\rho}(z)$ after taking into consideration the \emph{Swift}-BAT detection threshold.  They find that $\phi(z)$ decreases sharply above $L_{\rm cut} \sim 53$, and find no significant evidence for luminosity evolution as a function of redshift.  They also conclude that the GRB co-moving rate density follows the global star formation rate out to $z \sim 2-3$ and then flattens at higher redshift.  Based on there results, I adopt the following analytic descriptions for $\phi_{L}$ and $\dot{\rho}(z)$

\begin{equation} \label{phiz}
\phi_{L} = \frac{dN}{d~\rm{log}~L} = (L/L_{\rm cut})^{a_{L}}; L < L_{\rm cut} = (L/L_{\rm cut})^{b_{L}}
\end{equation}

\begin{eqnarray} \label{rhoz}
\dot{\rho}(z) = \frac{dN}{dz} \propto (1+z)^{g_{0}}; z < 0.97 \propto (1+z)^{g_{1}}; \nonumber\\
0.97 < z < z_{1} \propto (1+z)^{g_{2}}; z > z_{1}
\end{eqnarray}

where  $a_{L} = -0.22$ and $b_{L} = -2.89$ and ($g_{0}, g_{1}, g_{2}$) = (3.4, -0.3, -8) for $z_{1} = 4.5$ \citep{Butler10}.

\subsection{Converting Photons to Counts} \label{sec:Counts}

\begin{figure}
\includegraphics[height=.23\textheight,keepaspectratio=true]{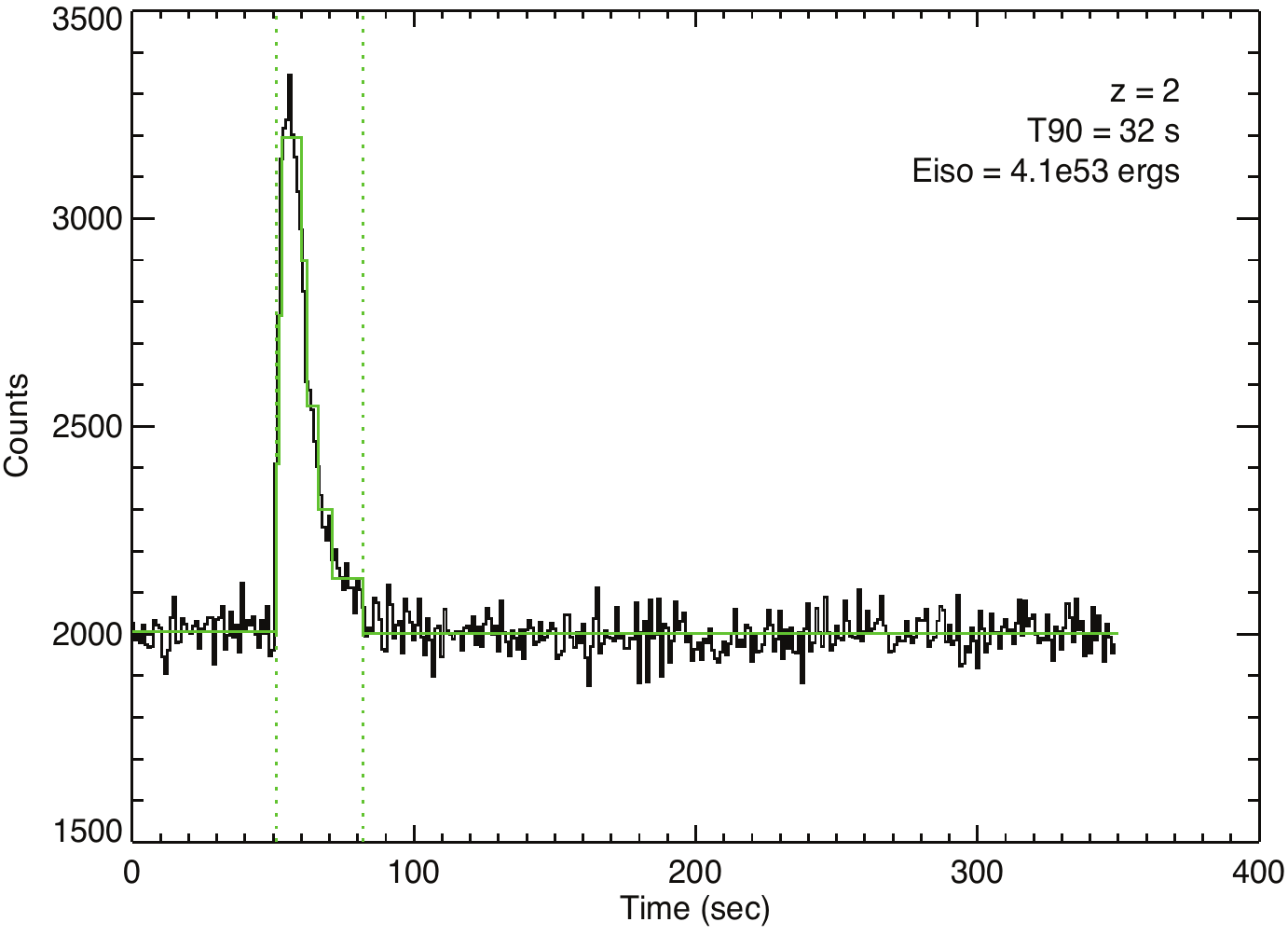}
\caption{The resulting count light curve obtained by folding the time-resolved photon spectra through a detector response and integrating the resulting individual time-resolved count spectra.  The burst  fluence and energetics are based on the duration obtained from the count light curve}
\label{Fig:LightCurves}
\end{figure}

Gamma-ray detectors trigger on counts accumulated by the instrument over a specified time interval.  The triggering algorithms used to measure the significance of these accumulated counts can be quite complicated, as is the case with the \emph{Swift}-BAT \citep{Gehrels04}, but was rather straight forward for the BATSE instrument \citep{Fishman94} onboard the \emph{Compton Gamma-ray Observatory}.  For BATSE, the instrument could trigger whenever there was an 5$\sigma$ excess of counts accumulated over three possible time timescales (64 ms, 256 ms, or 1024ms) compared to a running time averaged background computed every 17 seconds.

Whether a GRB triggers the instrument depends not only on the intrinsic luminosity of the burst and its distance from Earth, but also on the location of $E_{\rm pk,obs}$ in the detector's energy window.  In addition, a burst's spectral and temporal properties are inferred by observers using count data produced by the instrument, which likewise depends on $E_{\rm pk,obs}$.  Therefore, I convert the resulting photon model into a count model in order to asses the detectability of the burst as well as to measure the resulting burst properties as they would have been inferred by the observer.

In order to convert the observer frame photon model into a counts model, I take the simulated photon data cube and fold it through an instrument data response matrix (DRM).  For the purposes of this analysis, I use a BATSE response file that was generated for a real burst which occurred nearly at zenith for one of BATSE's Large Area Detectors (LAD).   The DRM describes the distribution of counts over the instrument's energy channels due to the arrival of photon of a given energy.  The result is a counts data cube as a function of channel energy, time, and count rate.  

I then add a Poisson distributed, energy dependent, background spectrum derived from the median backgrounds from a sample of BATSE detected bursts to each count spectrum.  A count light curve with a realistic background spectrum can then be produced as a function of time by integrating each individual time-resolved count spectra.  The resulting burst duration is then calculated using a Bayesian block algorithm.  An example of the resulting count light curve, for a  GRB placed at $z = 2$, and the subsequent Bayesian block reconstruction (solid green lines) can be found in Figure \ref{PhotonFluxLightCurve}.

Although only a small fraction of the GRBs that form the basis for the $E_{\rm pk} - E_{\rm iso}$ correlation were detected by BATSE, the broad energy range (20$-$1800 KeV) covered by the instrument and its relatively straight forward trigger criteria make it extremely easy to model for this analysis.  Furthermore, for purposes of the analysis presented here, I do not aim to reproduce the exact form of the observed $E_{\rm pk} - E_{\rm iso}$ correlation, but instead try to investigate how the combination of instrumental sensitivity and the cosmological nature of GRBs can work to produce an artificially strong correlation that does not exist in the source frame of the population. 

\begin{figure} 
\includegraphics[height=.23\textheight,keepaspectratio=true]{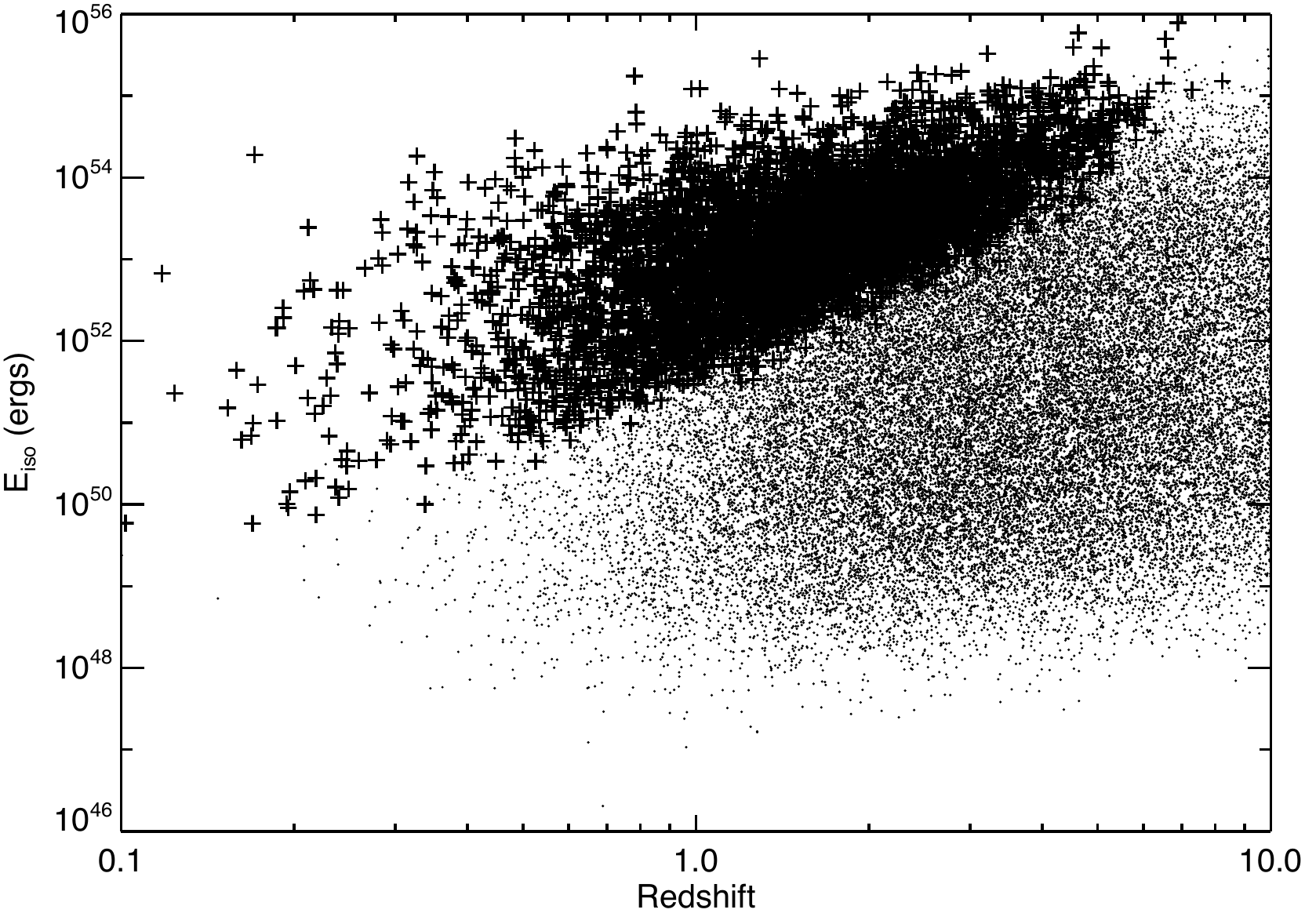}
\caption{$E_{\rm iso}$ data plotted versus redshift for our entire simulated sample.  GRBs with a trigger significance of more than 5.5 $\sigma$ above background in the 50-300 KeV energy range are shown as crosses whereas events that would have went undetected are shown as points.  A standard Malquest type bias produces the observed correlation between $E_{\rm iso}$ and redshift.}
\label{EisoVsRedshift}
\end{figure}

\section{Simulated Data Set and Analysis} \label{sec:Analysis}

Using the GRB pulse model and population synthesis code described in Sections 2.2 through 2.4, I simulate a set of 40,000 GRBs placed at a variety of redshifts following the co-moving rate density $\dot{\rho}(z)$ described in Equation \ref{rhoz}.  For each burst, the simulation code generates a time integrated PHA-I file by summing the counts data cube over a source duration as determined through the Bayesian block algorithm.  These PHA-I files, along with the original DRM used by the model, can then be read into XSPEC to obtain time-integrated spectral fits and photon and energy flux estimates in the 25-300 keV energy range, $F$ and $F_{\rm E}$ respectively.  

Using the XSPEC determined spectral parameters along with the measured energy flux and the Bayesian block determined duration, I can calculate the energy fluence $S_{\rm bol}$ integrated over an energy range of 10 to 10000 keV.  Using the recorded redshift, I can estimate the inferred isotropic equivalent energy $E_{\rm iso}$ and luminosity  $L_{\rm iso}$,  k-corrected to a standard energy range of 10 to 10000 keV, in the source frame. These inferred parameters are in addition to the \emph{true} bolometric $E_{\rm iso}$ and $L_{\rm iso}$, known to the simulation, but otherwise unknown to the observer. 

I find that although systematic differences do exist between these inferred energetics estimates and their true values due to instrumental effects, these differences do not heavily influence the results presented below.  I reserve a more detailed discussion of the systematic biases between a burst's inferred and true properties that are introduced by such instrumental biases for a future paper.  In this work, I will focus solely on the duration, $S_{\rm bol}$, $E_{\rm pk}$, $E_{\rm iso}$ and $L_{\rm iso}$ inferred by the user from the counts data for bursts that were detected by the instrument and use the true values to display the properties of bursts that were not detected by the instrument and hence for which no inferred values could be measured.

\section{Results} \label{sec:Results}

\begin{figure}
\includegraphics[height=.23\textheight,keepaspectratio=true]{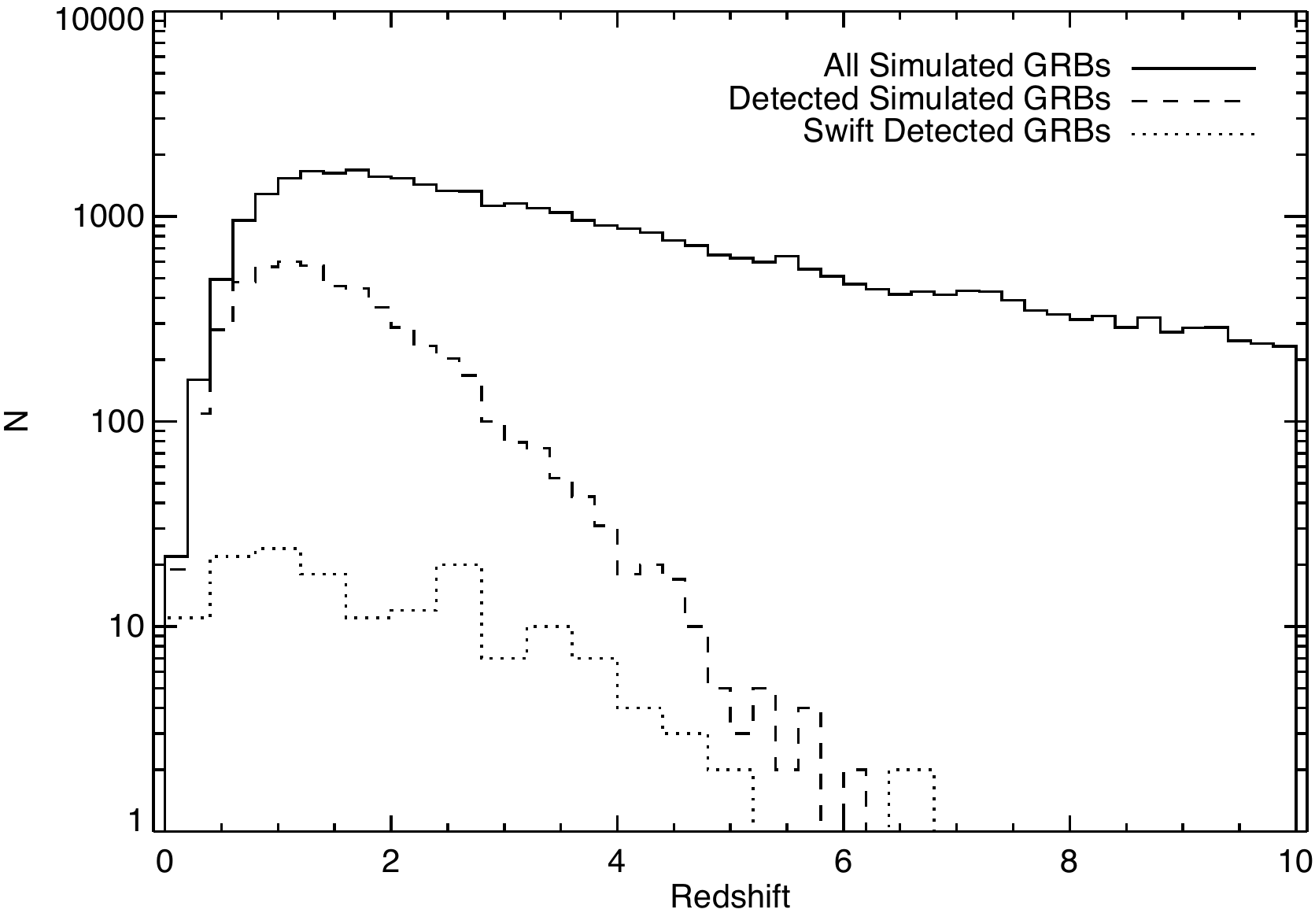}
\caption{The redshift distributions for all simulated (solid) and detected (dashed) bursts shown in comparison to the redshift distribution of \emph{Swift} detected bursts (dotted).}
\label{RedshiftDistribution}
\end{figure}

\begin{figure}
\includegraphics[height=.23\textheight,keepaspectratio=true]{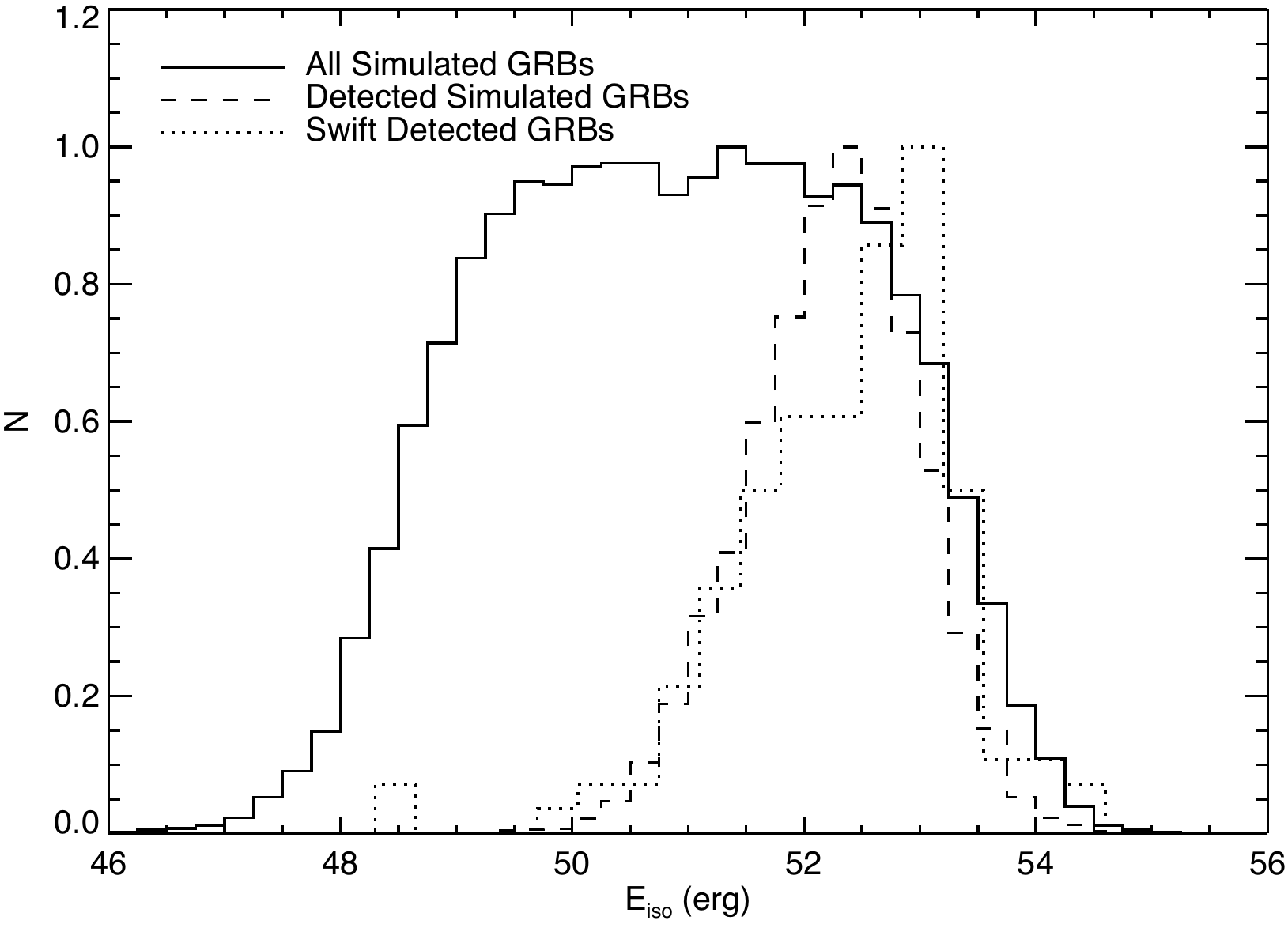}
\caption{The distribution of $E_{\rm iso}$ for all simulated (solid line) and detected (dashed) bursts in comparison with the \emph{Swift} determined $E_{\rm iso,\emph{Swift}}$ distribution.}
\label{EisoDistribution}
\end{figure}

\begin{figure} 
\includegraphics[height=.23\textheight,keepaspectratio=true]{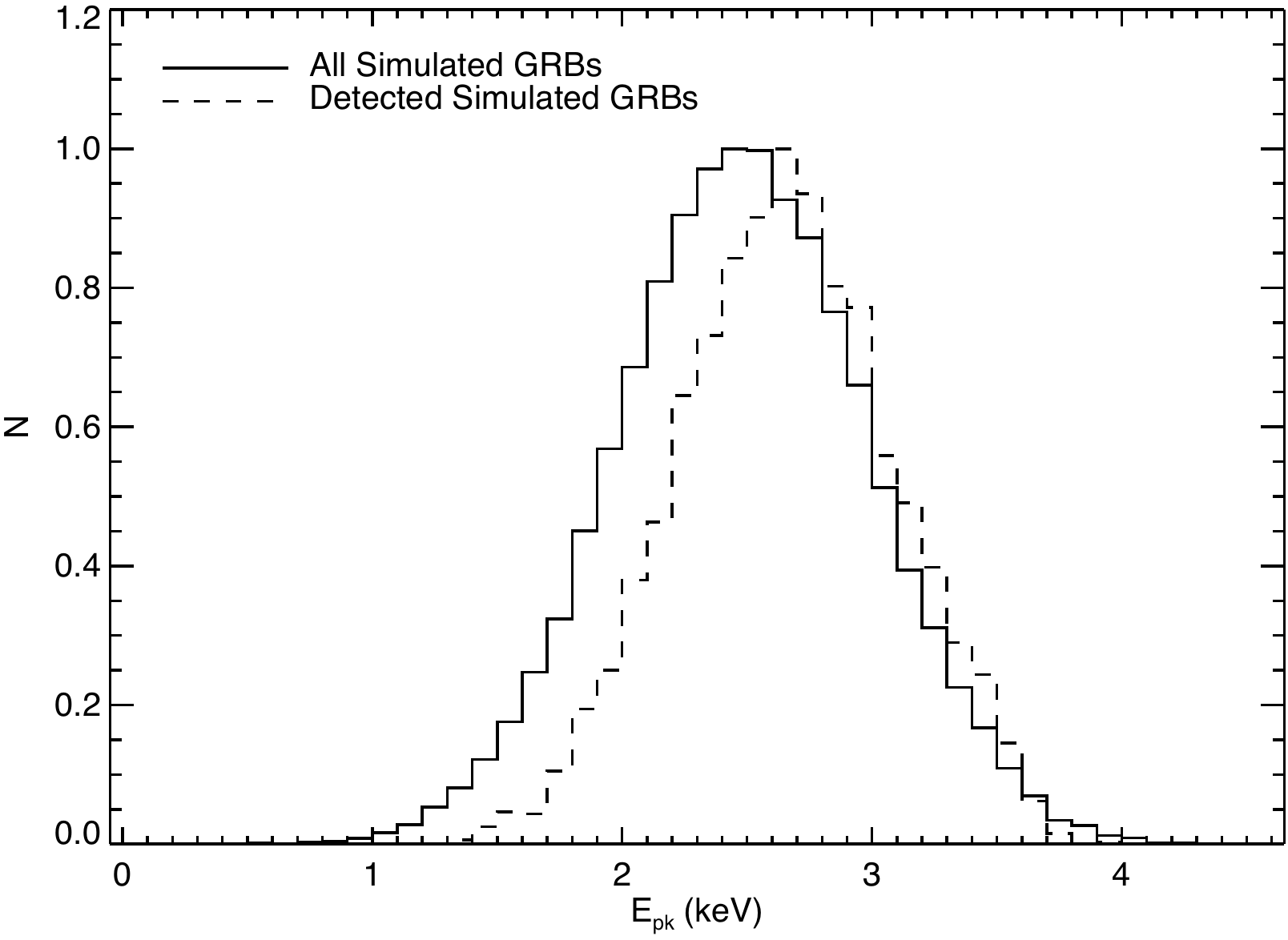}
\caption{The $E_{\rm pk,obs}$ distribution for all simulated bursts (solid) and detected (dashed) bursts.  The low end of the detected $E_{\rm pk,obs}$ distribution is influenced by the detector's limited sensitivity below $\sim 20$ keV, whereas the high end of the detected $E_{\rm pk,obs}$ distribution reflects the true cut-off in the simulated population.}
\label{EpkDistribution}
\end{figure}

The resulting $E_{\rm iso}$ data plotted versus redshift for our entire simulated sample can be seen in Figure \ref{EisoVsRedshift}.  GRBs with a count rate yielding a trigger significance of more than 5.5 $\sigma$ above background in the 50-300 KeV energy range are shown as crosses whereas events that would have went undetected are shown as points.  A standard Malquest type bias produces the observed correlation between $E_{\rm iso}$ and redshift, emphasizing the fact that luminous bursts are statistically found at higher redshifts.  The redshift distributions for all simulated bursts (solid) and detected bursts (dashed) are shown in figures \ref{RedshiftDistribution}, with the redshift distribution of \emph{Swift} detected bursts (dotted) also shown for comparison.  The decline in the observed GRB population at high redshift, compared to the simulated sample, is largely an effect of the sensitivity of the detector.  I find that without invoking luminosity evolution, the relative number of GRBs occurring at high redshift must increase in order to explain the number of high redshift detections.  This qualitative observation is in agreement with similar conclusions reported by \citet{Butler10}.

The distribution of $E_{\rm iso}$ for all simulated events (solid line) is shown in Figure \ref{EisoDistribution} along with the distribution of $E_{\rm iso}$ for the bursts that would have triggered the detector (dashed line).  The simulated population ranges from $10^{47} < E_{\rm iso} < 10^{55}$, but the resulting observed population drops off below $E_{\rm iso} < 10^{52}$ erg due to the detector threshold.  Conversely, the high end of the detected $E_{\rm iso}$ distribution is not influenced by the instrument, and instead reflects the cut-off that I assume for the GRB luminosity function.  The dotted line shows the \emph{Swift} $E_{\rm iso}$ distribution as determined by \citet{Butler10} for comparison.  Although slightly shifted to higher values, the \emph{Swift} observed $E_{\rm iso}$ distribution roughly matches our simulated distribution. 

The $E_{\rm pk,obs}$ distribution for all simulated bursts (solid line) and detected bursts (dashed line) is shown in Figure \ref{EpkDistribution}.  Similar to the case of the detected $E_{\rm iso}$ distribution, the low end of the detected $E_{\rm pk,obs}$ distribution is largely influenced by the limited sensitivity of the detector below $\sim 25$ keV, whereas the high end of the detected $E_{\rm pk,obs}$ distribution reflects the true cut-off in the simulated population.

\begin{figure}
\includegraphics[height=.23\textheight,keepaspectratio=true]{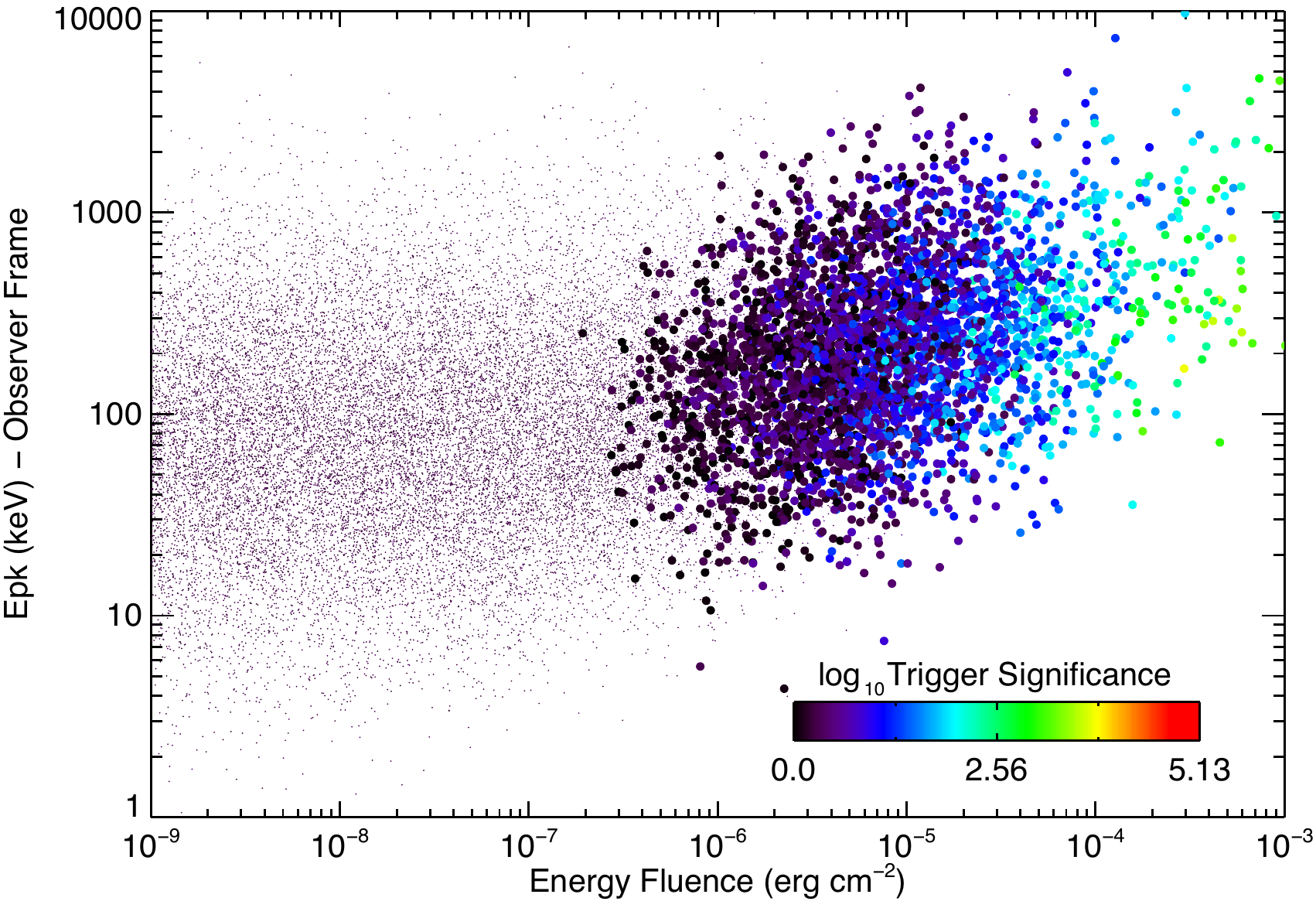}

\caption{The observer frame $E_{\rm pk,obs}$ versus the observed energy fluence $S_{\rm bol}$ for simulated (dots) and detected (circles) bursts.  The color gradient represents trigger significance of each burst in the 50-300 KeV energy range.  Increasing the detection threshold acts to narrow the observed population.}
\label{EpkVsFluenceVsTriggerSignificance}
\end{figure}

In Figure \ref{EpkVsFluenceVsTriggerSignificance}, I plot $E_{\rm pk,obs}$ versus the observed energy fluence $S_{\rm bol}$, integrated over an energy range of 10 to 10000 keV, for our entire simulated data set (dots), along with the bursts that would have triggered the instrument (circles).   The color of each data point represents the simulated burst's trigger significance in the 50-300 KeV energy range as seen by the detector, which can be seen to correlate with the observed energy fluence.  The boundary between the detected and undetected populations tracks the energy dependance of the detector's effective area, with the weakest detections occurring near the center of the detector's energy window, where the instruments sensitivity is maximized.  By following the color gradient, it can be seen that increasing the detection threshold acts to narrow the observed population.  The results presented in Figure \ref{EpkVsFluenceVsTriggerSignificance} are in qualitative agreement with similar modeling of the BATSE threshold as a function of energy found by \citet{Nava08}.  As discussed by \citet{Nava08}, I find that the lack of detected GRBs in the high $S_{\rm bol}$, low $E_{\rm pk,obs}$ regime of Figure \ref{EpkVsFluenceVsTriggerSignificance} is not due to any limitations of the detector, but rather starts to reflect the intrinsic cutoff of the GRB population as we run out of bright, nearby events.

\begin{figure}
\includegraphics[height=.23\textheight,keepaspectratio=true]{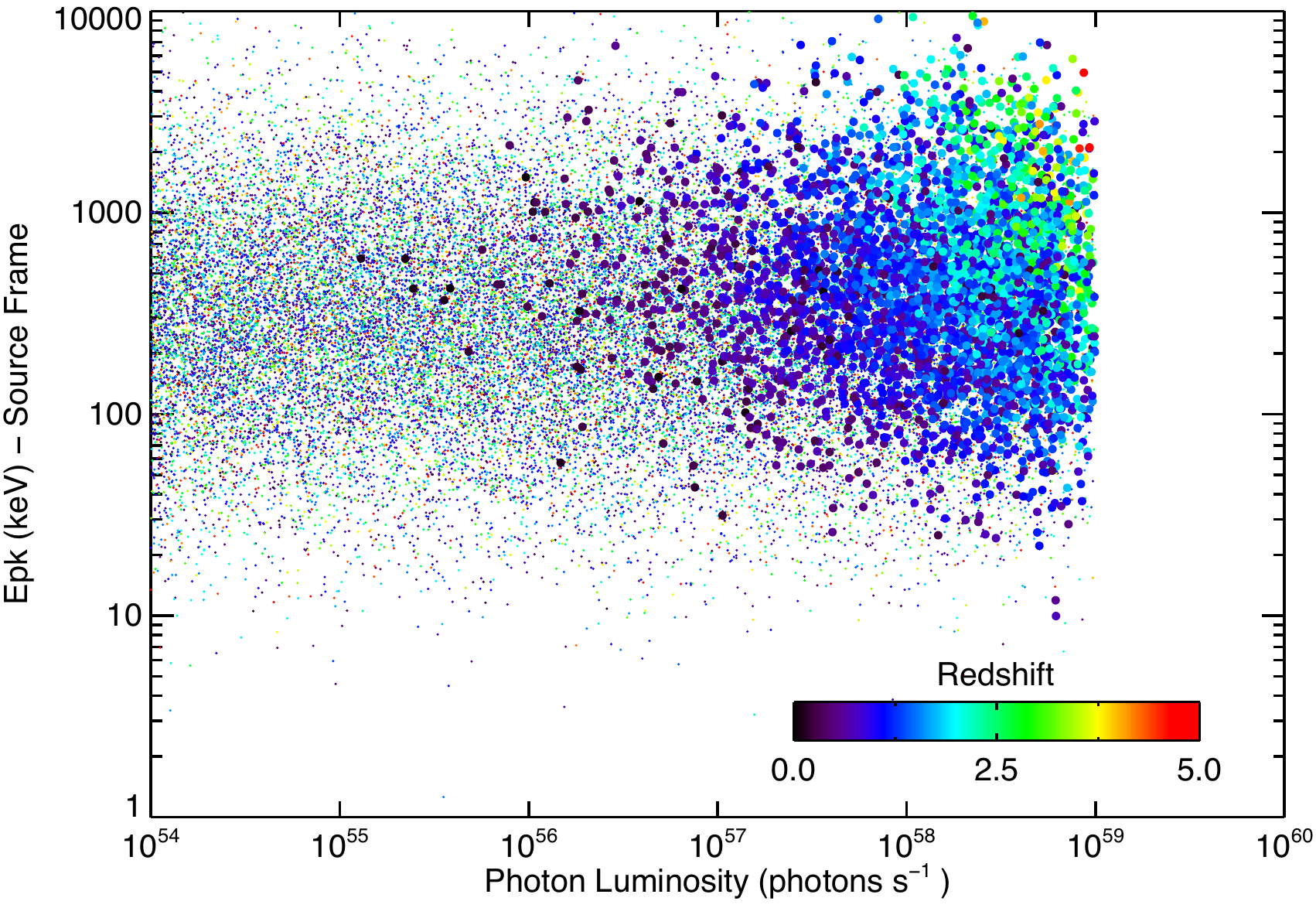}

\caption{Source frame $E_{\rm pk}$ versus the isotropic equivalent photon luminosity $L_{\rm iso}$ for all simulated (dots) and detected (circles) bursts.   The color of each data point represents the simulated burst's redshift.  Although no correlation is present between $L_{\rm iso}$, $E_{\rm pk}$, and redshift for the entire sample as a whole, a very distinct pattern emerges when considering only the detected events.}
\label{EpkTrueVsPhotonLuminosityVsRedshift}
\end{figure}

To understand the nature of this cutoff in the energy fluence distribution, we turn to Figure \ref{EpkTrueVsPhotonLuminosityVsRedshift}, where I plot the source frame $E_{\rm pk}$ versus the isotropic equivalent photon luminosity $L_{\rm iso}$ for our entire simulated data set (dots), along with the bursts would have triggered the instrument (circles).   The color of each data point now represents the simulated burst's redshift.  A broad scatter plot seen when considering the entire data set reflects the fact that I assume no intrinsic correlation between $L_{\rm iso}$, $E_{\rm pk}$, or redshift in the process of generating the simulated population.  The sharp drop in the number of events at the right end of the plot reflects the relatively steep luminosity function assumed in our population synthesis code, resulting in the cutoff in the $S_{\rm bol}$ distribution seen in Figure \ref{EpkVsFluenceVsTriggerSignificance}.

Several patterns become apparent when I consider only the detectable events, which form a triangular region in the $E_{\rm pk}$ versus $L_{\rm iso}$ plot.  Although no correlation is present between $L_{\rm iso}$, $E_{\rm pk}$, and redshift for the entire sample as a whole, a very distinct pattern emerges when considering only the detected events.  These patterns are due to the energy dependance of the instrument's detection threshold. The decreasing effective area near the high and low bounds of the instrument energy window becomes increasingly important for low luminosity bursts, decreasing the detection efficiency for low luminosity bursts with very low and very high $E_{\rm pk,obs}$ values.  On the other hand, extremely bright bursts are relatively easier to detect, even if $E_{\rm pk,obs}$ is outside of the detector's energy window, hence the wider range of detected bursts on the right side of Figure \ref{EpkTrueVsPhotonLuminosityVsRedshift}.  Finally, although intermediate luminosity bursts are seen at all redshifts, we only see the most luminous events at high redshift, again due to the Malquest bias.  More importantly, luminous bursts with low $E_{\rm pk}$ are not seen at very high redshift since their $E_{\rm pk,obs}$ values would be redshifted well outside of the detector's energy window, resulting in a low observed flux and making them less likely to be detected.  These two factors, a Malquest bias and the energy dependent detection threshold, produces a redshift gradient that is correlated with both $E_{\rm pk}$ and $L_{\rm iso}$. 

\begin{figure}
\includegraphics[height=.23\textheight,keepaspectratio=true]{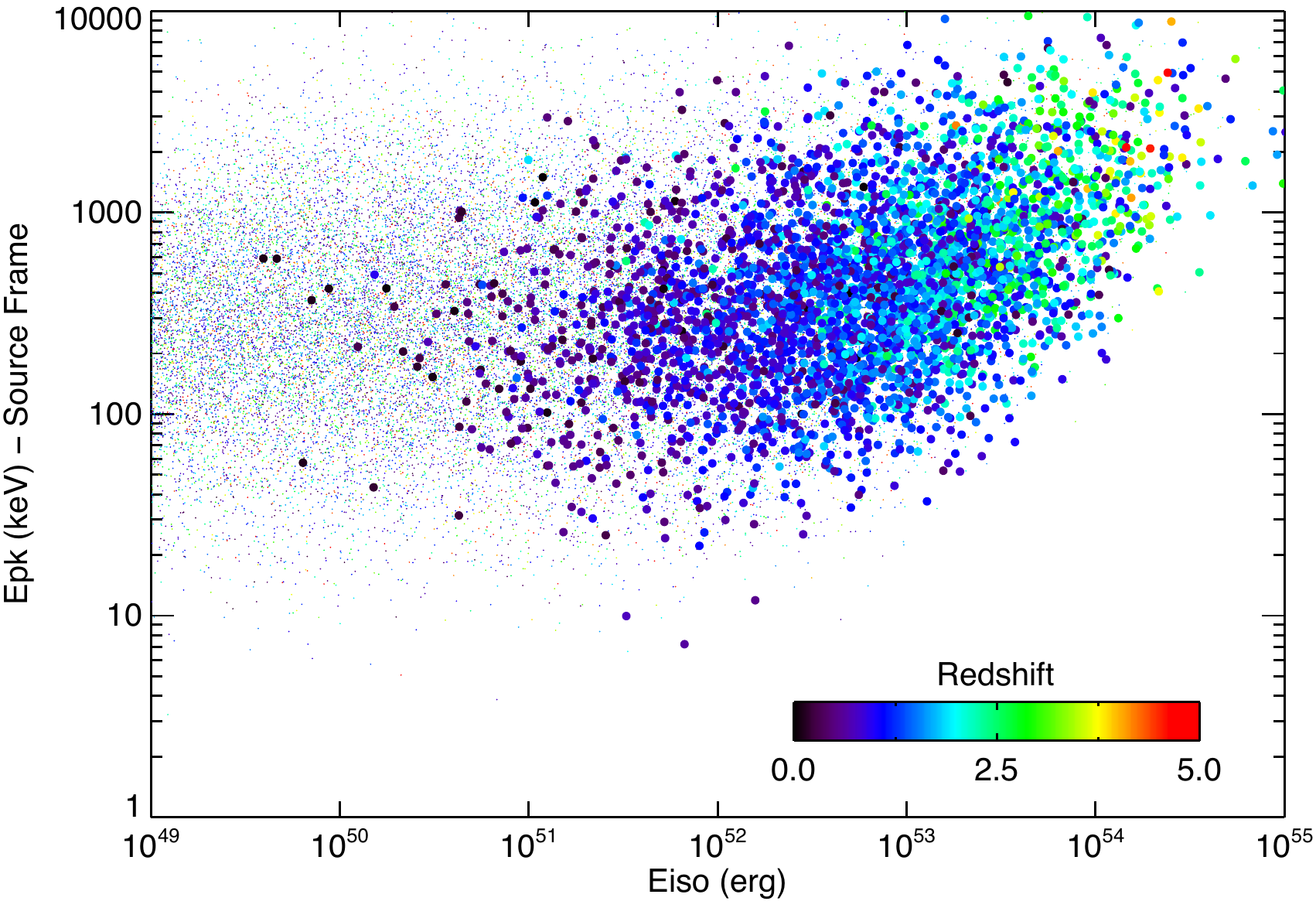}
\caption{Source frame $E_{\rm pk}$ versus $E_{\rm iso}$ for all simulated (dots) and detected (circles) bursts.   The color of each data point represents the simulated burst's redshift.  A broad positive correlation appears between $E_{\rm iso}$ and $E_{\rm pk}$ when considering only the detected population.}
\label{EpkTrueVsEisoBolometricTrueVsRedshift}
\end{figure}

In Figure \ref{EpkTrueVsEisoBolometricTrueVsRedshift}, I plot $E_{\rm iso}$ versus $E_{\rm pk}$ for our entire simulated data set (dots), along with the bursts that would have triggered the instrument (circles). The color of each data point again represents the burst's redshift.  Again, a broad scatter plot between the simulated $E_{\rm iso}$ and $E_{\rm pk}$ values for the entire data set (detected and undetected) reflects the fact that I assumed no intrinsic correlation between a GRB's $E_{\rm pk}$, photon luminosity $L_{\rm iso}$, and duration in the process of generating the simulated population.  Despite this, a broad positive correlation does appears between $E_{\rm iso}$ and $E_{\rm pk}$ among the GRBs that would have been detected by the instrument.  

\begin{figure}
\includegraphics[height=.23\textheight,keepaspectratio=true]{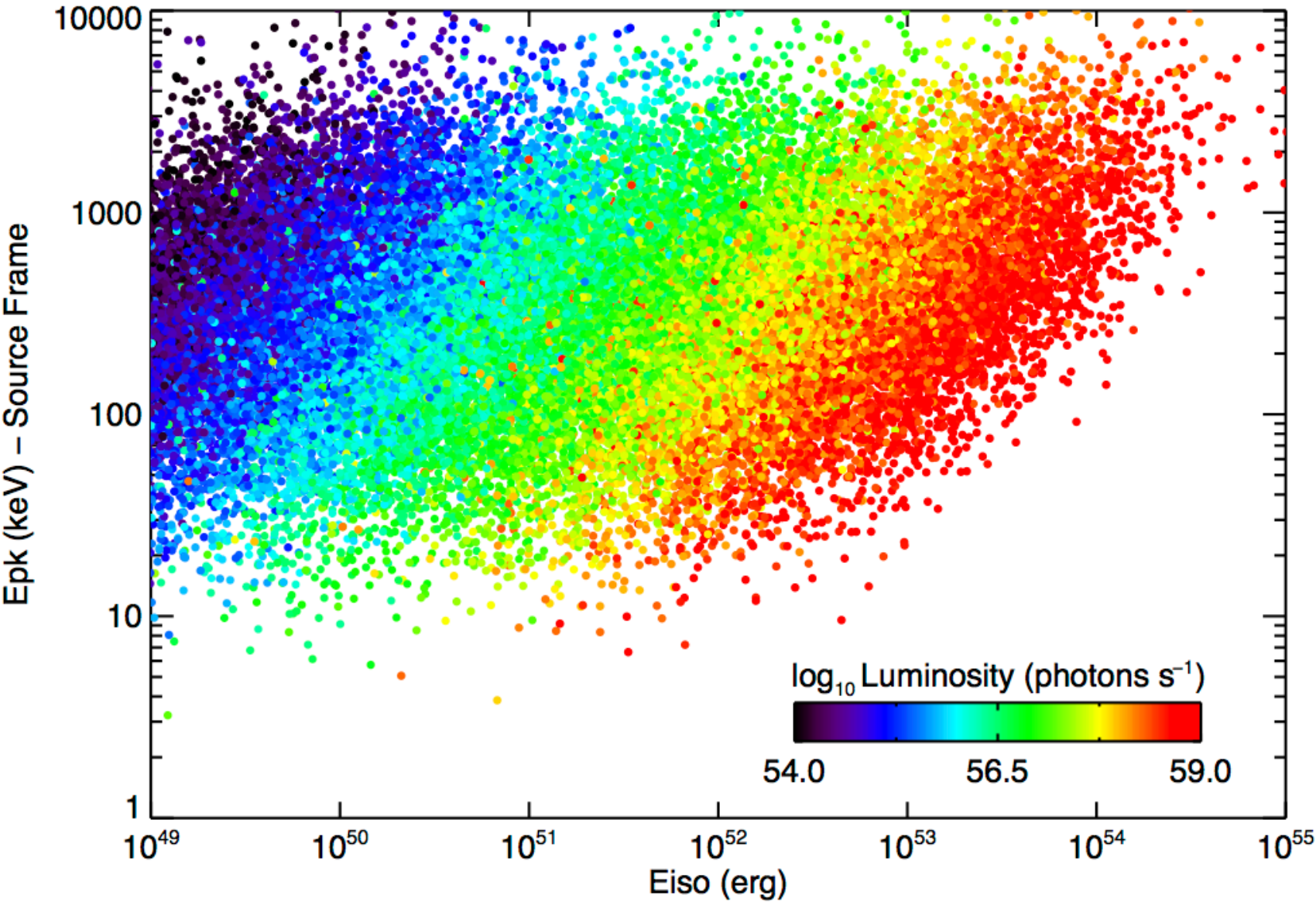}
\caption{Source frame $E_{\rm pk}$ versus $E_{\rm iso}$ with a color gradient representing $L_{\rm iso}$.  Areas of constant $L_{\rm iso}$ can be seen to track a linear correlation between $E_{\rm pk}$ and $E_{\rm iso}$ due to the conversion from photon luminosity to total energy, modulo the burst duration.}
\label{EpkTrueVsEisoBolometricTrueVsLuminosity}
\end{figure}

\begin{figure*}
\plottwo{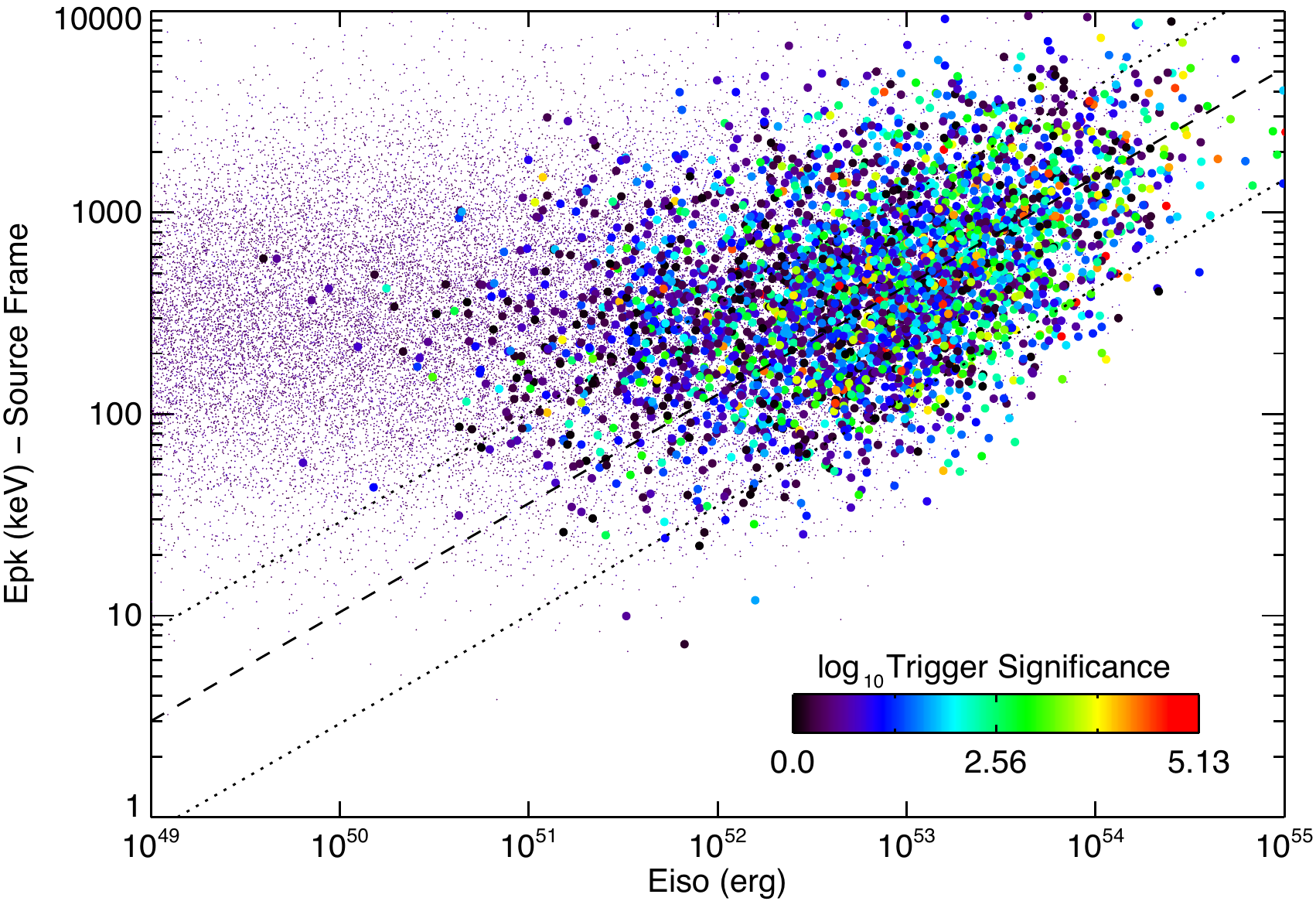}{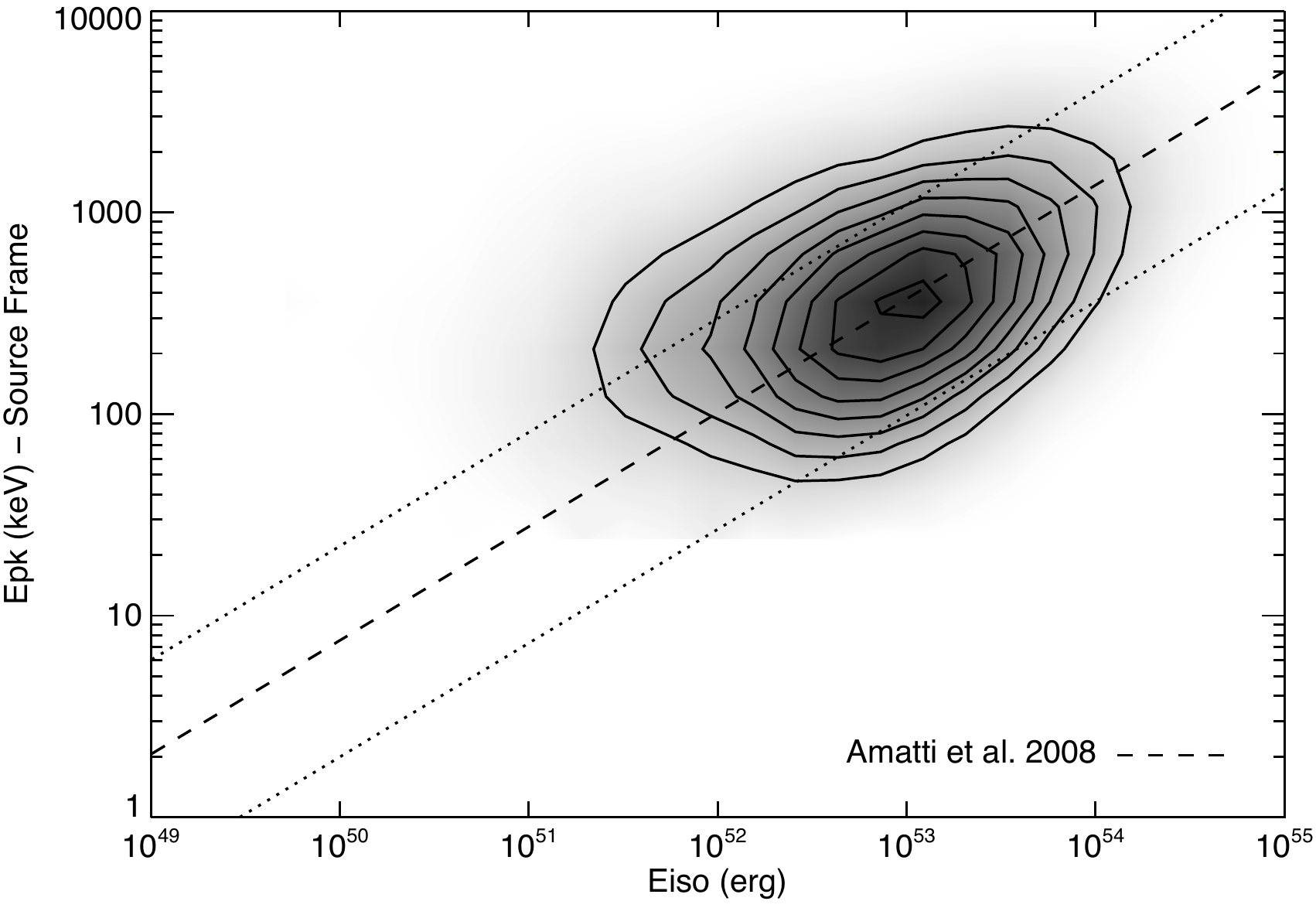}
\caption{{\emph Left Panel:} $E_{\rm iso}$ versus $E_{\rm pk}$ for all simulated (dots) and detected (circles) bursts.  The color of each point represents the burst's trigger significance in the 50-300 KeV for the detected population. The bursts with the highest trigger significance start to form a very distinct pattern.{\emph Right Panel:} The conversion of the left panel into a 2-dimensional probability distribution representing the most likely set of $E_{\rm iso}$ and $E_{\rm pk}$ values to be detected by the instrument once all of the various selection biases have been taken into account.  The dashed and dotted line represent the slope, normalization, and 1 sigma scatter of the $E_{\rm pk}-E_{\rm iso}$ correlation reported by \citet{Amati08}.}
 \label{EpkTrueVsEisoBolometricTrueVsSNRatioTrigger}
\end{figure*}

The slope of the lower-right edge of this correlation is roughly linear and can be understood as occurring by definition of the Band spectral model used in our simulations.  The Band function is comprised of two power-law components smoothly joined by an exponential.  $E_{\rm pk}$ is defined as the location of the break between these components and reflects the peak of the ${\nu}F_{\nu}$ spectra.  So, for GRBs with identical $L_{\rm iso}$, the events with higher $E_{\rm pk}$ values will by definition exhibit larger energy luminosities and hence larger $E_{\rm iso}$, modulo the GRB duration.  This can be seen more easily in Figure \ref{EpkTrueVsEisoBolometricTrueVsLuminosity}, where I plot $E_{\rm pk}$ versus $E_{\rm iso}$ with a color gradient representing $L_{\rm iso}$.  Areas of constant $L_{\rm iso}$ can be seen to track a linear correlation between $E_{\rm pk}$ and $E_{\rm iso}$.  So the lower-right edge that appears in Figure \ref{EpkTrueVsEisoBolometricTrueVsRedshift} reflects the sharp cutoff in our prescribed luminosity function above $L_{\rm cut} \sim 53$ transformed into an energy luminosity, which is inherently correlated with the burst's spectra.   
Likewise, the upper-left edge of the observed correlation reflects the shape of the detector threshold observed between $L_{\rm iso}$ and $E_{\rm pk}$ in Figure \ref{EpkTrueVsPhotonLuminosityVsRedshift}, but now transformed from photon luminosity to total energy, again modulo the GRB duration. 

As with the photon luminosity, there is a very distinct correlation between redshift, $E_{\rm iso}$, and $E_{\rm pk}$, with increasingly energetic and hard events only being detected at high redshift.  Burst's with low $E_{\rm iso}$ are only detectable at low redshift, with only a slight dependance on $E_{\rm pk}$ as described above.  Bursts with both high $E_{\rm iso}$ and high $E_{\rm pk}$ are rare and therefore the GRB luminosity function dictates that these events are most likely to occur at higher redshift, due to the increase in the observed volume.  Bursts with high $E_{\rm iso}$ and low $E_{\rm pk}$ are also rare, but become increasingly difficult to observe at high redshift because their $E_{\rm pk,obs}$ will be redshift below the detector's energy window.  In fact, low $E_{\rm pk}$ of all energies suffer the same selection bias, relegating low $E_{\rm pk}$ events to low and intermediate redshifts.  The net effect of these selection biases is a modification of the slope correlating $E_{\rm iso}$ and $E_{\rm pk}$ from the linear relationship expected from the nature of the Band function to a shallower slope due to the depletion of burst's with intermediate to high $E_{\rm iso}$ and low $E_{\rm pk}$ values. 

We can see this effect more easily in the left panel of Figure \ref{EpkTrueVsEisoBolometricTrueVsSNRatioTrigger} where I again plot $E_{\rm iso}$ versus $E_{\rm pk}$ for our entire simulated data set (dots), along with all bursts that would have triggered the instrument (circles).  The color coding now represents the trigger significance in the 50-300 KeV for the detected population.  Although there are weakly detected GRBs throughout the detected range of $E_{\rm iso}$ versus $E_{\rm pk}$, the bursts with the highest trigger significance start to form a very distinct pattern.  The edge of the luminosity function becomes less prominent through the depletion of low $E_{\rm pk}$ values with decreasing $E_{\rm iso}$, with the majority of bursts cluster at intermediate $E_{\rm iso}$ and $E_{\rm pk}$ values, producing a slope which is shallower than the linear relation between $E_{\rm iso}$, and $E_{\rm pk}$ for a given $L_{\rm iso}$.

I quantify the net effect these selection biases have on the observed GRB population by converting the left panel of Figure \ref{EpkTrueVsEisoBolometricTrueVsSNRatioTrigger} into the 2-dimensional probability distribution.  By mapping the intensity of this probability distribution, I can determine the GRB population most likely to be detected by the instrument.  The resulting probability distribution can be seen in the right panel of Figure \ref{EpkTrueVsEisoBolometricTrueVsSNRatioTrigger}.  

The intensity gradient and associated contours represent the most likely set of $E_{\rm iso}$ and $E_{\rm pk}$ values to be detected by the instrument once all of the various selection biases discussed above have been taken into account.  For comparison, I have plotted the slope, normalization, and scatter of the  $E_{\rm pk}-E_{\rm iso}$ correlation reported by \citet{Amati08}.  The slope of the reported correlation is roughly $\alpha \sim 0.54$, matching the slope of the intensity plot produced from our simulated data set.  The slope of the probability distribution is governed by the edge of the luminosity function at high $E_{\rm iso}$ and $E_{\rm pk}$, but deviates at low $E_{\rm pk}$ due to the depletion of these events because of cosmological redshift.   

\section{Discussion} \label{sec:Discussion}

The GRB model presented here is a very simplified representation of the complex nature of GRBs.  All of the bursts produced by our simulation are single pulsed events with Band function representations of their photon spectra.  Furthermore, the code explicitly assumes that their spectral evolution is dictated by simple relativistic kinematics with a pronounced hard to soft evolution.  In reality, GRBs are typically described by far more complex time histories, exhibit a wider range of spectral shapes, and at times do not follow the hard to soft evolution predicted by relativistic kinematics.  Still, these simulations serve to demonstrate how selection effects caused by a combination of instrumental sensitivity and the cosmological nature of GRBs can act to produce a strong correlation between observed properties.

As with all flux limited observations, our knowledge of astrophysical sources will be heavily influence by the properties of the brightest events in the sample, since they are the easiest to detect, even if they are not the most common.  The results presented in the left and right panels of Figure \ref{EpkTrueVsEisoBolometricTrueVsSNRatioTrigger} show that this is no different for GRBs, with the brightest bursts in the GRB luminosity function being the most commonly detected events.  I find that the edge of the GRB luminosity function plays an important role in producing the boundaries of the observed correlation between $E_{\rm pk}$ and $E_{\rm iso}$.  The fact that no instrument has to date detected a GRB with an $E_{\rm iso} > 10^{55}$ erg is due to a cutoff in the source population and cannot be explained by any selection effects.  Therefore, the upper right region of the observed $E_{\rm pk} - E_{\rm iso}$ diagram reflects the true cutoff in the GRB population.  Despite this, our simulations show that this cutoff, when expressed in energy space, cannot alone explain the slope of the $E_{\rm pk} - E_{\rm iso}$ relation.  Instead, the non-detection of low $E_{\rm pk}$, high $E_{\rm iso}$ GRBs at high redshifts acts to flatten the slope of the relationship.  By examining the density of detected bursts in the $E_{\rm pk} - E_{\rm iso}$ parameter space, the simulations reveal that the area in which GRBs are most likely to be detected roughly matches the $\alpha \sim 1/2$ slope reported by \citet{Amati08}.

Still there are several facets of the $E_{\rm pk} - E_{\rm iso}$ that are not addressed by these simulations.  Foremost, I do not address the nature of X-ray flashes (XRFs) and the adherence of at least three XRFs with $E_{\rm pk,obs} < 20$ KeV (XRF~020903, XRF~050416, and XRF~060218) to the correlation.  Events with such low $E_{\rm pk,obs}$ values were not simulated in our analysis because of BATSE's lack of energy coverage at those energies.  Furthermore, the instruments that detected these events, BeppoSAX for XRF~020903 and \emph{Swift}-BAT for XRF~050416, and XRF~060218, have unique response functions and triggering criteria which prevent me from quantifying the observed parameter space in which these events are most likely to occur.

Independent of detector consideration though, if XRFs are drawn from the same photon luminosity function as their higher energy GRB counterparts, as suggested by \citet{Lamb03} and \citet{Sakamoto08}, then low $E_{\rm pk}$ events would by definition have low $E_{\rm iso}$ for a given photon luminosity, modulo the burst duration.  An XRF with $E_{\rm pk} \sim$ a few keV will not posses the same energy luminosity as a burst with $E_{\rm pk} \sim 500$ KeV of the same photon luminosity.  Therefore, the cutoff in the photon luminosity function that is responsible for the production of the edge in the $E_{\rm pk} - E_{\rm iso}$ diagram would, by definition, preclude the existence of high $E_{\rm iso}$, low $E_{\rm pk}$ events unless XRFs are drawn from a separate population with a very different photon luminosity function compared to cosmological GRBs.

If XRFs, XRRs, and GRBs do represent a continuum of events drawn from the same population, then Figure \ref{EpkVsFluenceVsTriggerSignificance} would indicate that soft and weak XRFs should have an equal number of hard and weak counterparts with $E_{\rm pk} > 100$ keV.  Such events are not in fact reported by \citet{Sakamoto08}, where the authors compile a comprehensive list of XRFs and XRRs detected by \emph{Swift}-BAT down to $S_{\rm E} \sim 10^{-7}$ erg cm$^{-2}$ in the 15-150 keV energy range.  Hard events at these fluence levels are reported, though, by \citet{Butler07}, who utilize a novel Bayesian spectral fitting technique to estimate $E_{\rm pk,obs}$ near or beyond the upper edge of the \emph{Swift}-BAT energy window.  Therefore, the results presented in Figure \ref{EpkVsFluenceVsTriggerSignificance} which point to the existence of hard and soft events at low fluence levels are largely consistent with the results found by \citet{Butler07}.  If GRBs have a broad photon luminosity function and $E_{\rm pk}$ distribution and occur at a wide range of redshifts, then it is expected that the vast majority of bursts seen in the observer frame should occupy the low $S_{\rm bol}$ regime while covering a wide range of $E_{\rm pk}$ values.

Only a small fraction of the GRBs that form the basis for the $E_{\rm pk} - E_{\rm iso}$ correlation were detected by BATSE, but the instrument's broad energy range and its relatively straight forward trigger criteria make it extremely easy to model for this analysis.  Furthermore, the purposes of the analysis presented here is not necessarily to match the exact form of the $E_{\rm pk} - E_{\rm iso}$ correlation observed by other instruments, but rather to demonstrate how the combination of detector thresholds and the population distribution can work to produce artificially strong correlations that neither exist in the source frame of the population or are tied to the underlying physics of the sources in question.

\section{Conclusions} \label{sec:Conclusions}

The simulations present here paint a complex picture regarding the nature of the observed correlation between $E_{\rm iso} - E_{\rm pk}$.  By simulating a population of GRBs using an assumed luminosity function and co-moving rate density, I find that 

\indent $\bullet$ The $E_{\rm pk}$, $L_{\rm iso}$, and $E_{\rm iso}$ distributions among the detected population can appear strongly correlated to the observer despite the existence of only a weak and broad correlation in the original simulated population.  \\
\indent $\bullet$ The energy dependance of any detector's flux limited detection threshold acts to produce a correlation between the source frame $E_{\rm pk}$ and $E_{\rm iso}$ for low luminosity GRBs, producing the left boundary of the observed $E_{\rm pk} - E_{\rm iso}$ correlation. \\
\indent $\bullet$  Very luminous GRBs are found at higher redshifts than their low luminosity counterparts due to the standard Malquest bias, causing bursts in the low $E_{\rm pk}$, high $E_{\rm iso}$ regime to go undetected because their $E_{\rm pk}$ values would be redshifted to energies at which most gamma-ray detectors become less sensitive, producing the right boundary of the observed correlation and flatten the slope of the relationship. \\
\indent $\bullet$ The origin of the observed correlation is a complex combination of the instrument's detection threshold, the intrinsic cutoff in the GRB luminosity function, and the broad range of redshifts over which GRBs are detected.  \\

Although the GRB model presented here is a very simplified representation of the complex nature of GRBs, these simulations serve to demonstrate how selection effects caused by a combination of instrumental sensitivity and the cosmological nature of an astrophysical population can act to produce an artificially strong correlation between observed properties that does not exist in the source frame of the population. 

\acknowledgements

D.K. acknowledges financial support through the Fermi Guest Investigator program.  We thank Nathaniel Butler, Robert Preece, and Josh Bloom for insightful discussions. This work was supported in part by the NASA Fermi Guest Investigator program under NASA grant number NNX10AD13G and the U.S. Department of Energy contract to the SLAC National Accelerator Laboratory under no. DE-AC3-76SF00515.

\bibliography{ms}

\end{document}